\documentclass[aps,prl,twocolumn,showpacs,superscriptaddress]{revtex4} 
\usepackage{graphicx}
\usepackage{amsmath}
\usepackage{amssymb}
\usepackage{epstopdf}
\usepackage{amsfonts}
\usepackage{dcolumn}
\usepackage{bigints}
\usepackage{xcolor}
\usepackage{hyperref}
\usepackage[normalem]{ulem}
\begin{document}
\title{Undamped Rabi oscillations due to polaron-emitter hybrid states in non-linear photonic wave guide coupled to emitters }
\author{J. Talukdar}
\address{Homer L. Dodge Department of Physics and Astronomy,
  The University of Oklahoma,
  440 W. Brooks Street,
  Norman,
Oklahoma 73019, USA}
\address{Center for Quantum Research and Technology,
  The University of Oklahoma,
  440 W. Brooks Street,
  Norman,
Oklahoma 73019, USA}
\author{D. Blume}
\address{Homer L. Dodge Department of Physics and Astronomy,
  The University of Oklahoma,
  440 W. Brooks Street,
  Norman,
Oklahoma 73019, USA}
\address{Center for Quantum Research and Technology,
  The University of Oklahoma,
  440 W. Brooks Street,
  Norman,
Oklahoma 73019, USA}
\date{\today}

\begin{abstract}
The collective dynamics of two non-interacting two-level emitters, which are coupled to a
structured wave guide that supports two-photon bound states, is investigated. Tuning the energy 
of the two emitters such that they are in resonance with the two-photon bound state energy
band, we identify parameter regimes where 
the system displays fractional populations and essentially undamped Rabi oscillations.
The Rabi oscillations, which have no analog in the single-emitter dynamics,
are attributed to the existence of a collective polaron-like photonic
state that is induced by the emitter-photon coupling.
The full dynamics is reproduced by a two-state model,
in which the photonic polaron interacts with the state $|e,e,\text{vac} \rangle$ (two emitters in their excited state
and empty wave guide) through a Rabi coupling frequency  that depends on the emitter separation.
Our work demonstrates that emitter-photon coupling
can lead to an all-to-all momentum space interaction
between two-photon bound states and tunable non-Markovian dynamics,
opening up a new direction for emitter arrays coupled to a waveguide. 
\end{abstract}
\maketitle

Multi-level emitters coupled to
a radiation field in a periodic structure
are essential for delivering on the promises surrounding the second
quantum revolution.
Ongoing research is exploring a variety of platforms, including
nano-photonic lattices~\cite{ref_roy,ref_nano-shah,ref_palma,ref_wg_many-bound,ref_few-ph}, plasmonic wave guides~\cite{ref_plasmom}, and superconducting resonator arrays~\cite{ref_resonator,ref_sqbit}
coupled to atoms~\cite{ref_reitz,ref_yalla,ref_hood}, quantum dots~\cite{ref_qdot}, quantum solid-state defects~\cite{ref_barclay, ref_acs-lukin}, or superconducting qubits~\cite{ref_astafiev,ref_hoi,ref_mlynek,ref_mirhosseini,ref_sundaresan}.
Applications range from quantum information processing to quantum networking to quantum simulations~\cite{ref_pcw, ref_rydberg-pcw,ref_tudela,ref_stannigel,ref_facchi, ref_zheng,ref_shah_entang}.
Recent experimental milestones include the heralded creation of a single collective excitation in a chain of atoms coupled
to a waveguide~\cite{ref_single-collective-exc} and the demonstration of photon (anti-) bunching for weak atom-photon coupling 
by taking advantage of dissipation~\cite{ref_anti-bunching}.
Emitters coupled to a wave guide also constitute a promising platform with which to
study fundamental questions associated with open quantum systems, with the emitters playing the role of the
system and the wave guide or electromagnetic modes playing the role of the bath~\cite{ref_rabl_atom-field,ref_rabl_non-linear,ref_fractional-pop,ref_sherman,ref_cirac_2d_short,ref_tpohl}.

Building on the tremendous successes of cavity quantum electrodynamics (QED),
wave guide QED  plays a key role in a plethora
of quantum technologies~\cite{ref_photon-mat_cqed,ref_cqed-jon-simon}. The coupling of one or more excited multi-level emitters to
a continuum of electromagnetic modes  leads, in most cases, to irreversible correlated radiation
dynamics~\cite{ref_dicke,ref_cummings}.
Quite generally,
the strong transverse confinement in a waveguide 
speeds up the radiation dynamics compared to the free case~\cite{ref_onedimwaveguide}.
Moreover, the directionality of a one-dimensional waveguide facilitates the build-up of correlations
(or anti-correlations)
 between 
emitters that are separated by distances larger than the natural wave length
of the wave guide
leading to superradiance, subradiance, and entanglement generation~\cite{ref_sinha, ref_chiral-zoller, ref_Haakh, ref_chaos, ref_flat, ref_goban, ref_subr,ref_bello,ref_poddubny, ref_molmer3, ref_molmer4}. The emergence of these 
characteristics can be explained in terms of constructive and destructive interferences. 
This work predicts long-lived oscillatory radiation dynamics for a generic waveguide QED set-up
that can be realized experimentally with existing state-of-the-art technology. The oscillatory radiation dynamics 
is distinct from the typically observed irreversible correlated radiation dynamics.

We consider a structured or non-trivial bath, namely a
wave guide with non-linearity that supports a
band of two-photon bound states (or more generally, a band of bound bath quantum pairs)~\cite{ref_rabl_non-linear}.
Working in the quantum regime, where the system contains just two excitations,
the influence of the non-trivial mode structure of the bath on the radiation dynamics is investigated
within a full quantum mechanical framework. 
Non-Markovian dynamics is observed. Rather counterintuitively, a regime is identified
where the radiation dynamics is described nearly perfectly by a two-state Rabi model.
An analytical framework that elucidates the underlying physical mechanism is developed.
It is shown that two emitters separated by multiple lattice sites are, in certain parameter
regimes, glued together and coupled to a wave guide with all-to-all momentum space interactions.
It is as if the band of two-photon bound states was feeling a localized (in real space) impurity that 
leads to the formation of a photonic polaron-like state with which the two-emitter unit interacts,
creating hybridized symmetric and anti-symmetric states that exchange population, undergoing essentially
undamped Rabi oscillations.

\begin{figure}[t]
\vspace*{-1.2cm}
\includegraphics[width=0.58\textwidth]{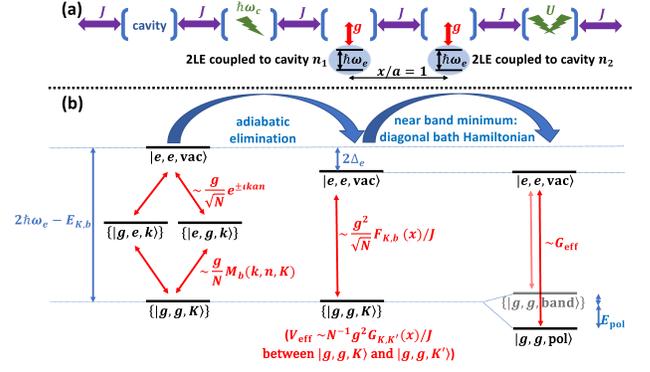}
\vspace*{-2.4cm}
\caption{(a) 
Schematic of the Hamiltonian $\hat{H}$.
The cavity array and two-level emitters (2LE) are shown; the role of the 
different energy terms is illustrated.
(b) Illustration of the Hilbert space structure of
$\hat{H}$ (left), $\hat{H}^{\text{adia}}$ (middle), and
$\hat{H}^{\text{2-st.}}$ (right). The matrix element $M_b(k,n,K)$ is defined in Ref.~\cite{ref_footnote-M}. 
Note that the energy difference $2 \hbar \omega_e-E_{K,b}$, Stark shift $2 \Delta_e$, and polaron energy $E_{\text{pol}}$ are not shown to scale.
}
\label{fig_schematic}
\end{figure}

Figure~\ref{fig_schematic}(a) illustrates 
the set-up.
The 
total Hamiltonian $\hat{H}$ consists
of
the system, 
tight-binding bath or wave guide,
and
system-bath 
Hamiltonians 
$\hat{H}_s$,
$\hat{H}_b$, and
$\hat{H}_{sb}$~\cite{ref_rabl_non-linear}, 
\begin{eqnarray}
\hat{H}_s=
\frac{\hbar \omega_e}{2}
\sum_{j=1}^{N_e} 
(\hat{\sigma}_{j}^z + \hat{I}_j),
\end{eqnarray}
\begin{eqnarray}
\label{eq_bath_ham}
\hat{H}_b =&&
\hbar \omega_c  \sum_{n=1}^{N} 
 \hat{a}_n^{\dagger} \hat{a}_n 
-J  \sum_{n=1}^{N} \left(
\hat{a}_n^{\dagger} \hat{a}_{n+1} + \hat{a}_{n+1}^{\dagger} \hat{a}_{n} 
\right) \nonumber \\
&&+\frac{U}{2} 
\sum_{n=1}^{N}
 \hat{a}_n^{\dagger}\hat{a}_n^{\dagger}\hat{a}_n\hat{a}_n ,
\end{eqnarray}
and
\begin{eqnarray}
\hat{H}_{sb}= 
g \sum_{j=1}^{N_e}
\left(
\hat{a}_{n_j} \hat{\sigma}_j^+ + \hat{a}_{n_j}^{\dagger}  \hat{\sigma}_j^-
\right),
\end{eqnarray}
 where $ \hbar \omega_e$,
 $\hbar \omega_c$, $J$, and $U$ denote the
 energy difference of the excited and ground state of the emitter, the
 photon energy in the middle of the single-photon band, the hopping energy, and
 the engineered or intrinsic onsite energy, respectively.
Since the coupling energy $g$ is small compared to $|U|$ and $J$,
counterrotating terms are  not included in $\hat{H}_{sb}$;
throughout, positive $g$ and $J$ and negative $U$ 
are considered
(positive $U$ yield the same results). 
The emitter operators  $\hat{\sigma}_{j}^z=|e \rangle _j \langle e| - |g \rangle _j \langle g|$,
$\hat{I}_j=|e \rangle _j \langle e| + |g \rangle _j \langle g|$,
$\hat{\sigma}_{j}^+ = |e\rangle _j \langle g|$,
and
$\hat{\sigma}_{j}^- = |g\rangle _j \langle e|$ act on the $j$th emitter 
located at lattice site $n_j$
with ground and excited states
$| g \rangle_j$ and $|e \rangle_j$. 
The bath operators $\hat{a}_{n_j}^{\dagger}$ and $\hat{a}_{n_j}$ create and destroy
a photon at lattice site $n_j$
($j=1,\cdots,N_e$ and $n_j \in 1,\cdots,N$).
Throughout, we consider $N_e=2$ emitters with
separation $x$, $x=n_1-n_2$, and large number
of lattice sites $N$.
The bath Hamiltonian $\hat{H}_b$ supports, due to the Kerr-like nonlinearity $U$,
a band of two-photon bound states, one
bound state with energy $E_{K,b}$ for each two-photon center-of-mass wave vector $K$~\cite{ref_boundstate_1991,ref_boundstate_1992,ref_molmer1, ref_molmer2,ref_valiente, ref_petrosyan}. The existence of these bound states has been confirmed
experimentally in photonic and cold atom optical lattice systems~\cite{nature2013_boundstate,winkler2006}. 
Throughout, the emitter energy 
is tuned such that $2 \hbar \omega_e$ is equal to
$E_{K^{(0)},b}$ at the uncoupled resonance wave vector $K^{(0)}$.
Since we are interested in the two-excitation subspace with $K^{(0)} a$ close to zero,
 the detuning $\delta$ is measured from the bottom
of the two-photon bound state band, $\delta = 2 \hbar \omega_e- 2 \hbar \omega_c + \sqrt{U^2 + 16J^2} $.

To describe the time evolution of the initial state
$|e,e, \text{vac} \rangle$,
we expand the time-dependent wave packet 
$|\Psi(t) \rangle$ as~\cite{ref_rabl_non-linear}
\begin{eqnarray}
\label{eq_wavepacket_ansatz}
|\Psi(t) \rangle  = 
&&
\exp(- 2\imath\omega_e t)
\bigg[
c_{ee}
|e,e,\mbox{vac} \rangle 
+\sum_K 
c_{K,b}
|g,g,K \rangle 
\nonumber \\
&&+ \sum_{k}  
c_{1 k}
|e,g,k \rangle + 
\sum_k 
c_{2 k}
|g,e,k \rangle
\bigg]
, 
\end{eqnarray}
where 
$c_{ee}(t)$, $c_{K,b}(t)$, $c_{1 k}(t)$, and $c_{2 k}(t)$
denote expansion coefficients, and
$|k\rangle=\hat{a}_k^{\dagger}|\text{vac}\rangle$ and $|K\rangle=\hat{P}_{K,b}^{\dagger}|\text{vac} \rangle$
single-photon states with wave vector $k$ and photon-pair states 
with center-of-mass wave vector $K$, respectively.
The operators $\hat{a}_k^{\dagger}$ and $\hat{a}_n^{\dagger}$ are related
via a Fourier transform. Our ansatz does not account for the two-photon scattering continuum
since it plays a negligible role for the parameter
combinations considered in this paper~\cite{footnote2}.

\begin{figure}[t]
\includegraphics[width=0.41\textwidth]{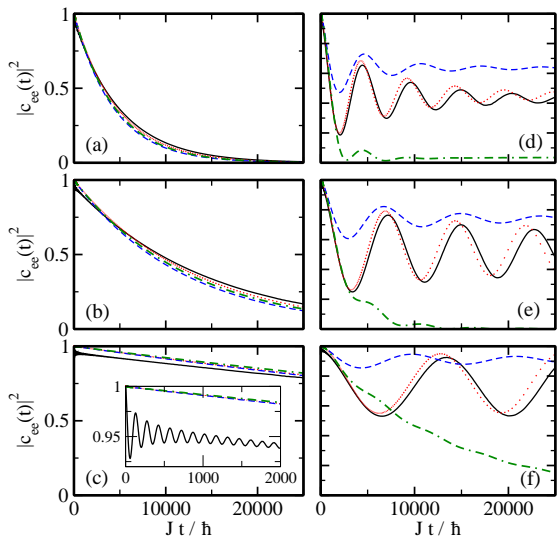}
\caption{$|c_{ee}(t)|^2$ as a function of $Jt/ \hbar$ for the
initial state $|e,e,\text{vac}\rangle$, $U/J=-1$, $g/J=1/50$, and
$\delta/J=0.0431$ (left) and $\delta/J=0.0011$ (right). Top, middle, and bottom rows are for $x/a= 0$, $5$, and $10$, respectively. Black solid, red dotted, blue dashed, and green dash-dash-dotted lines are obtained using $\hat{H}$, $\hat{H}^{\text{adia}}$, $\hat{H}^{\text{adia}}$ with $G_{K,K'}=0$, and $\hat{H}^{\text{adia}}$ with $G_{K,K'}=\Delta_e=0$, respectively.}
\label{fig_population}
\end{figure}

The  solid lines in the left column of Fig.~\ref{fig_population}  show
the population $|c_{ee}(t)|^2$ of the state $|e,e,\text{vac} \rangle$
as a function of time for 
$U/J=-1$, $g/J=1/50$,
$\delta/J=0.0431$, and  
$x/a=0$, $5$, and $10$,
obtained by propagating the ansatz given in Eq.~(\ref{eq_wavepacket_ansatz})
using
$\hat{H}$.
For this detuning, $|c_{ee}(t)|^2$
decreases approximately exponentially. This is the Markovian regime, discussed in Ref.~\cite{ref_rabl_non-linear}, where propagation with the adiabatic Hamiltonian $\hat{H}^{\text{adia}}$ 
yields quite accurate results (dotted,  dashed, and  dash-dash-dotted lines show results for three different variants of $\hat{H}^{\text{adia}}$). 
The adiabatic Hamiltonian $\hat{H}^{\text{adia}}$, which lives in a reduced Hilbert space that excludes the single-photon states $|e,g,k\rangle$ and $|g,e,k\rangle$, is introduced below [middle of Fig.~\ref{fig_schematic}(b)]. 
The inset of Fig.~\ref{fig_population}(c) 
for $x/a=10$ shows that the short-time behavior of $|c_{ee}(t)|^2$ deviates from a pure exponential 
decay.
This is due
to the fact that the dynamics is,
for $x/a \gg 1$, seeded by the creation of two uncorrelated photons. 
For larger times, the fall-off is, as for smaller separations, again governed by correlated two-photon dynamics.

When the emitter energy is set such that 
$|\delta|$
is very small ($K^{(0)} a \approx 0$), the radiation dynamics changes drastically. The right column of Fig.~\ref{fig_population} shows an example for $\delta/J=0.0011$.
For $x=0$ [Fig.~\ref{fig_population}(d)], the propagation under $\hat{H}$ (solid line) yields damped oscillatory behavior. In the long-time limit, the system is characterized by fractional steady-state atomic populations. This is analogous to the single-emitter case~\cite{ref_fractional-pop,ref_sherman}, where the emitter frequency is in resonance with the
single-photon scattering band.
In the single-emitter case, the term fractional steady-state atomic population is used to indicate that the system is
in a quasi-stationary state, which has appreciable overlap with the state $|e,\text{vac}\rangle$ and
the states $|g,k\rangle$~\cite{ref_fractional-pop}.
By analogy, we use the term fractional  steady-state atomic population  in our two-emitter case 
to indicate that the system is
in a quasi-stationary state, which has appreciable overlap with the state $|e,e,\text{vac}\rangle$ and
the states $|g,g,K\rangle$.
As the separation increases [Figs.~\ref{fig_population}(e)-\ref{fig_population}(f) show results for $x/a=5$ and $10$, respectively], the dynamics for the Hamiltonian $\hat{H}$ (solid lines) are characterized by slower oscillations and weaker  damping. For $x/a=10$, 
the oscillations   resemble nearly perfect two-state Rabi oscillations. 
Even though the emitters are coupled
to a bath, dephasing is essentially absent for large separations. These undamped
Rabi oscillations have no analog in the single-emitter system~\cite{ref_fractional-pop,ref_sherman}.

\begin{figure}[t]
\includegraphics[width=0.43\textwidth]{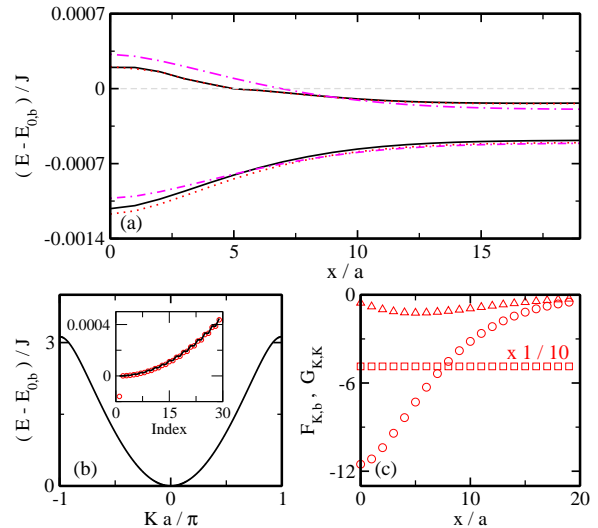}
\caption{Static results ($U/J=-1$, $g/J=1/50$, and $\delta/J=0.0011$).  (a) Black solid, red dotted, and magenta dash-dotted lines show the eigenenergies corresponding to hybridized states of $\hat{H}$, $\hat{H}^{\text{adia}}$, and $\hat{H}^{\text{2-st.}}$, respectively,  as a function of  $x/a$. The gray dashed line shows $(E-E_{0,b})/J=0$. (b) The black solid line shows  $E_{K,b}$ as functions of $Ka/\pi$ (main panel) and the state index (inset).  The red circles show the eigenenergies supported by $\hat{H}^{\text{adia}}_b$ (index 1 corresponds to the 
polaron-like
state). (c) The squares, circles, and triangles show the dimensionless quantities $\text{Re}[G_{K^{(0)},K^{(0)}}(x)]/10$,
$\text{Re}[F_{K^{(0)},b}(x)]$,  and $\text{Im}[F_{K^{(0)},b}(x)]$ as a function of $x/a$ for $K^{(0)}a / \pi =0.0152$. 
}
\label{fig_spect_band_int}
\end{figure}

\begin{figure}[t]
\includegraphics[width=0.43\textwidth]{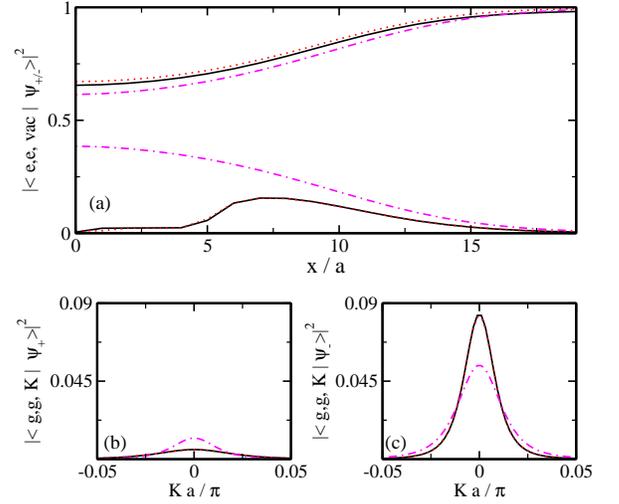}
\caption{State composition of hybridized polaron-emitter states 
($U/J=-1$,  $g/J=1/50$, and $\delta/J=0.0011$).
(a) Projection of $|e,e,\text{vac}\rangle$ onto 
$\Psi_+$
(upper three lines) and 
$\Psi_-$ (lower three lines)
as a function of $x/a$. 
Black solid, red dotted, and magenta dash-dotted lines are
obtained using $\hat{H}$, $\hat{H}^{\text{adia}}$, and $\hat{H}^{\text{2-st.}}$,
respectively.
(b)/(c) Projection of $\Psi_+$ and $\Psi_-$ onto $|g,g,K\rangle$ as a function of $Ka/\pi$ for $x/a=10$.
The line styles are the same as in (a); black solid and red dotted lines are nearly indistinguishable on the scale shown.
}
\label{fig_atom-pol_compo}
\end{figure}

The oscillation frequencies in Figs.~\ref{fig_population}(d)-\ref{fig_population}(f) correspond to the energy difference between the two energy eigenstates of $\hat{H}$ that have the largest overlap with $|e,e,\text{vac}\rangle$ [solid lines in  Fig.~\ref{fig_spect_band_int}(a)]; we label these states $\Psi_+$ and $\Psi_-$. For $x/a \gtrsim 5$, $\Psi_{\pm}$ have an energy that is smaller than $E_{K=0,b}$, i.e., both states are bound with respect to the $g=0$ two-photon bound state band [solid
line in Fig.~\ref{fig_spect_band_int}(b)]. For $x/a \lesssim 5$, the
energy of $\Psi_+$ remains below the bottom of the two-photon band while that of $\Psi_-$ lies in the continuum.
The quantity $|\langle e,e,\text{vac}| \Psi_+ \rangle|^2$ increases from about $0.66$ to $0.99$ as $x/a$ increases from $0$ to $20$ [upper solid line in Fig.~\ref{fig_atom-pol_compo}(a)];
$| \langle e,e,\text{vac}|\Psi_- \rangle|^2$, in contrast, is comparatively small for $x/a \lesssim 4$, increases for $x/a=5$ to $7$, and then slowly decreases as $x/a$ increases further
[lower  solid  line in Fig.~\ref{fig_atom-pol_compo}(a)].

To understand the emergence of the bound states and their  dependence on $x$, 
we adiabatically eliminate the states $|e,g,k\rangle$
and $|g,e,k\rangle$, i.e., we assume that the change of
the expansion 
coefficients $c_{1k}(t)$ and $c_{2k}(t)$ in Eq.~(\ref{eq_wavepacket_ansatz})
with time can be neglected~\cite{ref_rabl_non-linear}.
This
introduces a Stark shift $2 \Delta_e$
as well as 
effective 
momentum space interactions, proportional to $N^{-1}g^2 {G}_{K,K'}(x)/J$,
between 
two-photon bound states   with wave vectors $K$ and $K'$. 
Since the two-photon bound state   with wave vector $K$ is coupled to two-photon bound states with other $K'$, i.e., $G_{K,K'}(x)$ is non-diagonal, we refer to the effective interaction 
$N^{-1}g^2 {G}_{K,K'}(x)/J$
as an effective all-to-all momentum space interaction.
The spread of $G_{K,K'}(x)$ over a wide range of center-of-mass wave vectors is discussed in detail in
Ref.~\cite{long-paper}; it plays a critical role when the absolute value of the
detuning $\delta$ is small.  
The structures of $\hat{H}$ and the Hamiltonian $\hat{H}^{\text{adia}}$ after adiabatic elimination are sketched, respectively, in the left and middle diagrams of Fig.~\ref{fig_schematic}(b).
For the larger $\delta$ considered in Fig.~\ref{fig_population} (left column), the $2 \Delta_e$ and $G_{K,K'}(x)$ terms have negligible effects 
on the radiation dynamics [the dotted, dashed, and dash-dash-dotted lines 
in Figs.~\ref{fig_population}(a)-\ref{fig_population}(c) agree well]; as a consequence, Ref.~\cite{ref_rabl_non-linear} set them to zero in their reduced Hilbert
space description. 
For the smaller $\delta$ (right column of Fig.~\ref{fig_population}), in contrast, both terms have
a non-negligible effect on the dynamics as evidenced by the fact that the dotted, dashed, and dash-dash-dotted lines in Figs.~\ref{fig_population}(d)-\ref{fig_population}(f)
disagree.

The adiabatic Hamiltonian $\hat{H}^{\text{adia}}$ 
contains the system, 
bath,
and system-bath Hamiltonians
$\hat{H}^{\text{adia}}_{s}$, 
$\hat{H}^{\text{adia}}_{b}$, and
$\hat{H}^{\text{adia}}_{sb}$,
\begin{eqnarray}
\hat{H}_{s}^{\text{adia}}= 2 \Delta_e |e,e,\text{vac} \rangle \langle e,e,\text{vac}|,
\end{eqnarray}
\begin{eqnarray}
\hat{H}_b^{\text{adia}} &=& 
\sum_K 
 E_{K,b}
|g,g,K \rangle \langle g,g,K|  \nonumber \\
&&+\sum_{K,K'}  \frac{g^2}{J N} G_{K,K'}(x) |g,g,K \rangle \langle g,g,K' | ,
\end{eqnarray}
and
\begin{eqnarray}
\hat{H}^{\text{adia}}_{sb} = \sum_K \frac{g^2}{J\sqrt{N}} F_{K,b}(x)|g,g,K\rangle \langle e,e,\text{vac} | + h.c.
\end{eqnarray}
The analytical expressions for 
the effective interactions $g^2 N^{-1/2}  F_{K,b}(x)/J$
and  $ g^2 N^{-1} G_{K,K'}(x)/J$ are lengthy and not reproduced here~\cite{ref_rabl_non-linear,long-paper}.
The dotted lines in Figs.~\ref{fig_population}, \ref{fig_spect_band_int}(a), and \ref{fig_atom-pol_compo}(a) show the results obtained by propagating the initial state
$|e,e,\text{vac} \rangle$
with 
$\hat{H}^{\text{adia}}$. The dotted lines agree quite well with the full calculation (solid lines)
for all detunings and separations considered, suggesting that the reduced Hilbert space model
captures the key physics. Thus, we use it 
to develop physical intuition. 

To start with, we analyze
the
$K \approx K' \approx K^{(0)} \approx 0$ portion of $\hat{H}^{\text{adia}}_b$, which 
should govern the radiation dynamics when
$|\delta/J|$ approaches zero. In this regime,
the imaginary part of $G_{K,K'}(x)$ is vanishingly small.
In fact, since $G_{K,K'}(x)$ is (excluding real overall factors) a sum over products 
$[M_b(k,n,K)]^*[M_b(k,n,K')]$, it is purely real for $K=K'$; here,
$M_b(k,n,K)$ measures the overlap between $|K\rangle$ and $\hat{a}_n^{\dagger}|k\rangle$~\cite{long-paper}. 
When $K$ and $K'$ differ,
$G_{K,K'}(x)$ can be loosely thought of as an autocorrelation function for
the overlaps. 
Importantly,
 the real part, shown for $\delta/J= 0.0011$ by the squares in Fig.~\ref{fig_spect_band_int}(c), is negative and nearly independent of $x$.
Considering that the states $|e,g,k\rangle$ and $|g,e,k\rangle$ that are being eliminated
adiabatically contain information on the emitter locations,
it is remarkable that  $\text{Re}[G_{K,K'}(x)]$ is nearly independent of the emitter separation $x/a$. 
The behavior of $G_{K,K'}(x)$ is discussed in detail in Ref.~\cite{long-paper}.
If we replace $E_{K,b}$ by $E_{0,b}$ (i.e., use a flat band)
and $G_{K,K'}(x)$ by $G_{K^{(0)},K^{(0)}}(x)$, then the eigenenergies of the 
bath Hamiltonian are $E_{0,b}-(N-1) g^2 N^{-1} G_{K^{(0)},K^{(0)}}(x)/J$ (one-fold degenerate) and
$E_{0,b}+g^2  N^{-1}G_{K^{(0)},K^{(0)}}(x)/J$ [$(N-1)$-fold degenerate].
The eigenstate of the one-fold degenerate bound state reads $N^{-1/2}\sum_K |K\rangle$.
This bound state can be interpreted as a bosonic quasi-particle that lives in the
Hilbert space of the dressed infinite
cavity array, with the dressing coming from the effective photon-pair--photon-pair interactions 
that are introduced by the adiabatic elimination. 
Since the eigenstate of the bosonic quasi-particle in the cavity array Hilbert space can be written as a superposition of 
$|K \rangle$ states, we refer to it as a polaron-like state.

While the flat band model overestimates the binding energy of the 
polaron-like bound state
by a fair bit,
 it shows that the attractive all-to-all interactions $ g^2 N^{-1}  G_{K,K'}(x)/J$ are responsible for the
 fact that the band of bound photon pairs splits into a collective polaron-like bound state and a band that is slightly shifted
 upward compared to the $G_{K,K'}(x)=0$ case.
 This interpretation continues to hold when a more accurate treatment is employed.
The band curvature can be thought of as introducing
a wave vector cutoff $(L_{\text{eff}})^{-1}$.
Taylor-expanding $E_{K,b}$ up to order $(Ka)^2$,
making the ansatz $|\text{pol}\rangle = \sum_K d_K |K\rangle$
with $d_K=2 N^{-1/2}  (L_{\text{eff}}^{-1}a/2)^{3/2}  / [(Ka)^2 +(L_{\text{eff}}^{-1}a/2)^2]$, 
and treating 
$L_{\text{eff}}$ as a variational parameter, the energy
$E_{\text{pol}}$
of the polaron $|\text{pol} \rangle$ can be found analytically.
For the parameters considered in Fig.~\ref{fig_spect_band_int}(b),
the analytical result is in excellent agreement with the lowest eigenenergy of $\hat{H}_b^{\text{adia}}$, which is shown in Fig.~\ref{fig_spect_band_int}(b)
by the  circle for state index 1.

Since $G_{K^{(0)},K^{(0)}}(x)$ is, for fixed $\delta/J$ and $U/J$, 
approximately
independent of $x$, the separation dependence displayed in Figs.~\ref{fig_population}(d)-\ref{fig_population}(f) must enter through $F_{K^{(0)},b}(x)$.
Figure~\ref{fig_spect_band_int}(c) shows that $\text{Re}[F_{K^{(0)},b}(x)[$ (circles) has a strong $x$ dependence and is much larger, in magnitude, than $\text{Im}[F_{K^{(0)},b}(x)]$ (triangles).
Throughout, we work with parameter combinations where
the resonant wave number $K^{(0)}$  is much smaller than $a$, implying that the 
oscillatory behavior of $F_{K^{(0)},b}(x)$, encoded in 
 $\sin(Ka)$ and $\cos(Ka)$ terms, does not play a role~\cite{long-paper}.
 This is in contrast to earlier studies where the emitter was in resonance with 
 the single-photon band and where the oscillatory nature of the coherent and dissipative dipole-dipole interactions  
played a role (see, e.g., Ref.~\cite{ref_rabl_atom-field}).
Rewriting $\hat{H}^{\text{adia}}$ in the basis in which $\hat{H}_b^{\text{adia}}$ is diagonal, we find that the state $|e,e,\text{vac} \rangle$ couples comparatively strongly to the state $|g,g,\text{pol}\rangle$ and comparatively weakly to all other bath states. The dynamics in the $|\delta/J| \rightarrow 0$ limit is thus approximately described by the two-state Hamiltonian
$\hat{H}^{\text{2-st.}}$,
\begin{eqnarray}
\label{eq_ham_two_state}
\hat{H}^{\text{2-st.}} = \hat{H}_s^{\text{adia}}
+ E_{\text{pol}} |g,g,\text{pol}\rangle \langle g,g,\text{pol}|
+ \nonumber \\
(G_{\text{eff}} 
|g,g,\text{pol} \rangle \langle e,e,\text{vac}|  
+ h.c.
).
\end{eqnarray}
Using our variational expression for $|g,g,\text{pol} \rangle$,
we find
\begin{eqnarray}
\label{eq_geff}
G_{\text{eff}} = \frac{g^3(U^2+16J^2)^{1/4}}{2J^{5/2}}F_{K^{(0)},b}(x)|G_{K^{(0)},K^{(0)}}(x)|^{1/2}.
\end{eqnarray}
The eigenenergies of 
the hybridized 
polaron-emitter states $\Psi_+$ and $\Psi_-$
supported by Eq.~(\ref{eq_ham_two_state})
for $U/J=-1$, $g/J=1/50$, and $\delta/J=0.0011$
[dash-dotted lines in Figs.~\ref{fig_spect_band_int}(a)]
agree reasonably well with those of $\hat{H}$ when $x/a$ is large. State $\Psi_+$ is symmetric (the coefficients of $|e,e,\text{vac}\rangle$ and $|g,g,\text{pol}\rangle$
are both positive) while $\Psi_-$ is anti-symmetric (the coefficients have opposite signs).

The two-state description deteriorates with decreasing separation; the state composition of the more weakly bound state $\Psi_-$, which 
has a smaller overlap with the emitter state $|e,e,\text{vac}\rangle$ 
[lower three lines in Fig.~\ref{fig_atom-pol_compo}(a); Fig.~\ref{fig_atom-pol_compo}(c)] than the more deeply
bound state $\Psi_+$ [upper three lines in Fig.~\ref{fig_atom-pol_compo}(a); Fig.~\ref{fig_atom-pol_compo}(b)],
deviates notably from 
that
obtained by diagonalizing $\hat{H}$. 
In fact, for $x/a \lesssim 5$, the first excited state of
$\hat{H}$ is no longer  a simple superposition of
$|e,e,\text{vac} \rangle$ and $|g,g,\text{pol}\rangle$ but instead contains  multiple nearly degenerate energy eigenstates with energy close to $E_{K=0,b}$. In the dynamics, this results in dephasing, thereby explaining the damping observed in Figs.~\ref{fig_population}(d)-\ref{fig_population}(e).
We emphasize that the emergence of the three different regimes
(exponential decay, fractional populations, and Rabi oscillations), illustrated  
in Fig.~\ref{fig_population} for the separations of $x/a=0$, $5$, and $10$,
depends on the values of $U/J$, $g/J$, and $\delta/J$.
For the same $U/J$ and $\delta/J$, the Rabi oscillation regime
can be understood by analyzing the interplay between $E_{\text{pol}}$ (which contains a term that scales as $-g^4/J^4$), 
$G_{\text{eff}}$ (which is proportional to $g^3/J^3$), and $\Delta_e$ (which is proportional to $g^2/J^2$)
within the two-state Hamiltonian $\hat{H}^{\text{2-st.}}$~\cite{long-paper}.

In summary, our analysis shows that the essentially undamped Rabi oscillations are associated with
population exchange between two hybridized polaron-emitter states. 
These states are distinct from previously predicted hybridized states~\cite{ref_rabl_atom-field,ref_manybody, ref_optexp, ref_lonigro, ref_stewart, ref_javadi}.
For the parameters considered in this paper,
the more weakly-bound hybridized state merges into the continuum for $x/a \lesssim 5$, making the emergence of long-lived Rabi oscillations an intriguing emitter separation-dependent
long-range phenomenon. When the emitters are close together, 
the radiation dynamics, starting with $|e,e,\text{vac}\rangle$ at $t=0$, leads to quasi-stationary fractional populations.
When the emitters are spaced further apart, regular revivals are observed.
We emphasize the crucial role of the
Stark shift $2 \Delta_e$ and the attractive all-to-all momentum space interactions. 
Neglecting these terms yields the dash-dotted lines in Figs.~\ref{fig_population}(d)-\ref{fig_population}(f). Setting $2\Delta_e$ to the correct value but using $G_{K,K'}(x)=0$ yields the dashed lines.

Our work illustrates that the structure of the bath Hamiltonian 
with Kerr-like non-linearity can be modified non-trivially---introducing attractive all-to-all momentum space interactions---through the coupling to two two-level emitters, resulting in qualitatively new radiation dynamics. 
Continuing to work in the two-excitation manifold, extension to arrays of regularly spaced emitters where neighboring emitters have a fixed separation (simple emitter lattice) or alternating separations (emitter superlattice) offers the prospect of establishing non-trivial bath-induced correlations between separated emitter pairs. 
Taking an alternative viewpoint, this work points toward utilizing 
emitters
to create bath Hamiltonian with unique characteristics.
Our analysis assumes that losses from the wave guide can be neglected. Over the time scales considered, this should be justified for several state-of-the-art experiments.

{\em{Acknowledgement:}}
Support by the National Science Foundation through grant number PHY-1806259 is gratefully acknowledged. This work used the OU Supercomputing Center for Education and Research (OSCER) at the University of Oklahoma (OU). A discussion with Kieran Mullen on impurities in solid state systems is  gratefully acknowledged.

\end{document}